\documentclass[prd,aps,preprint,tightenlines,showpacs,nofootinbib,superscriptaddress]{revtex4-1}
\usepackage{mathrsfs}
\usepackage{amsfonts}
\usepackage{amsmath}
\usepackage{array}
\usepackage{verbatim}
\usepackage{bm}
\usepackage{epsfig}
\usepackage{graphicx,color}
\usepackage{relsize}
\usepackage{lineno}
\usepackage{float}
\usepackage{multirow}
\RequirePackage{xspace}
\def\reg{{I\!\!R}}
\def\pom{{I\!\!P}}

\begin{document}

\title{\boldmath Near  threshold heavy vector meson photoproduction \\
 at LHC and EicC}

\author{Ya-Ping Xie}
\email{xieyaping@impcas.ac.cn}
\affiliation{Institute of Modern Physics, Chinese Academy of Sciences,
	Lanzhou 730000, China}
\affiliation{University of Chinese Academy of Sciences, Beijing 100049, China}

\author{V.~P. Gon\c{c}alves}
\email{barros@ufpel.edu.br}
\affiliation{High and Medium Energy Group, \\
	Instituto de F\'{\i}sica e Matem\'atica, Universidade Federal de Pelotas\\
	Caixa Postal 354, CEP 96010-900, Pelotas, RS, Brazil}
\affiliation{Institute of Modern Physics, Chinese Academy of Sciences,
	Lanzhou 730000, China}

\begin{abstract}
The exclusive $J/\Psi$ and $\Upsilon$ photoproduction in fixed - target collisions at the LHC and $ep(A)$ collisions at the Electron ion collider in China (EicC) is investigated considering different models for the treatment of the vector meson production at low energies, close to the threshold. Results for the total cross sections and associated distributions are presented. We predict a large number of $J/\Psi$ events at the LHC in the rapidity range covered by the LHCb detector. For the EicC, our predictions point out that a detailed analysis of the near threshold $J/\Psi$ and $\Upsilon$ photoproduction is feasible. Moreover, our results indicate that the modeling of the near threshold vector meson production can be constrained by future experimental analyzes at the LHC and EicC.
\end{abstract}

\maketitle

\section{Introduction}
\begin{figure}[t]
	\centering
	\begin{tabular}{cc}
	\includegraphics[width=0.45\textwidth]{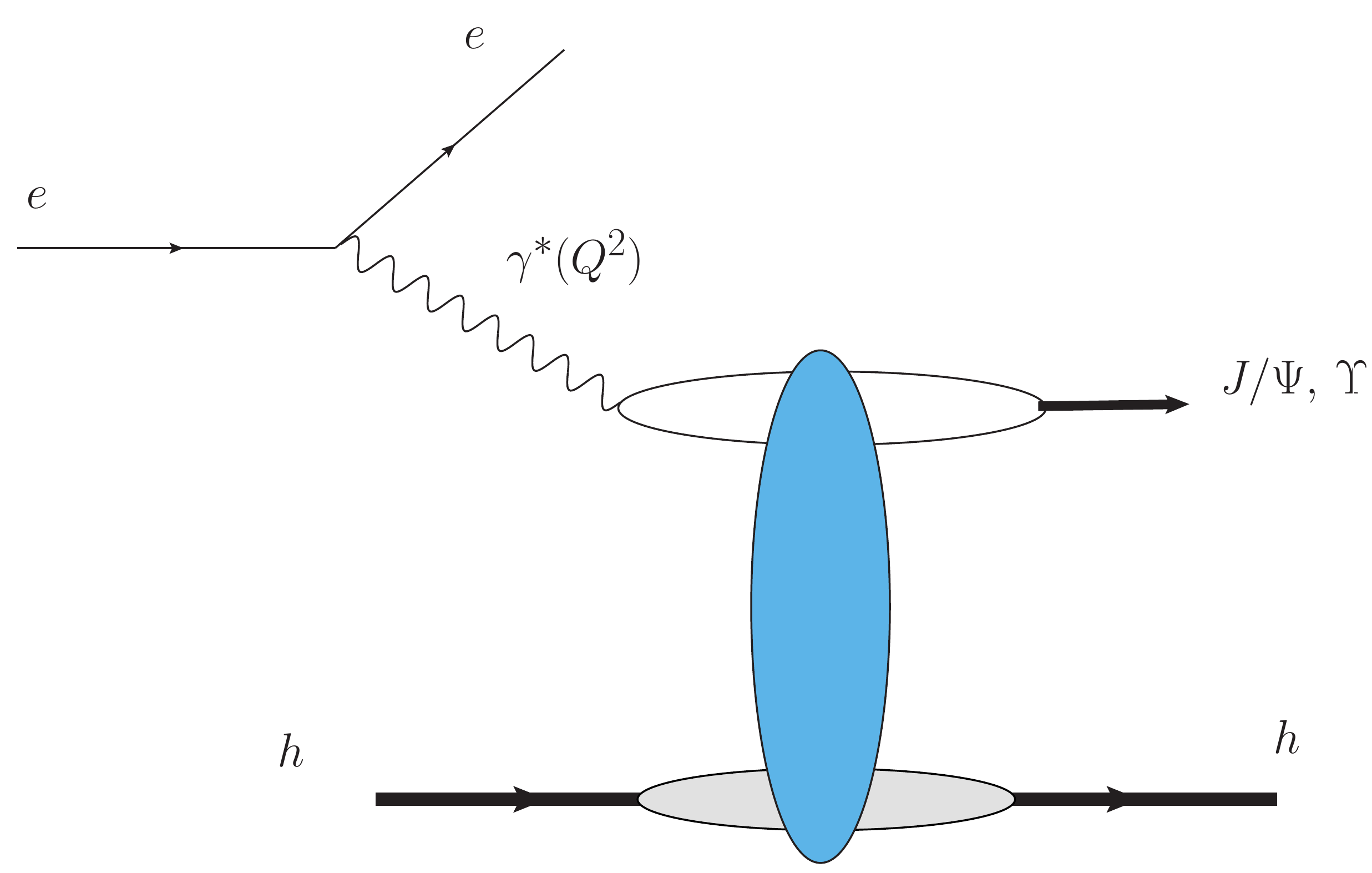} &
		\includegraphics[width=0.5\textwidth]{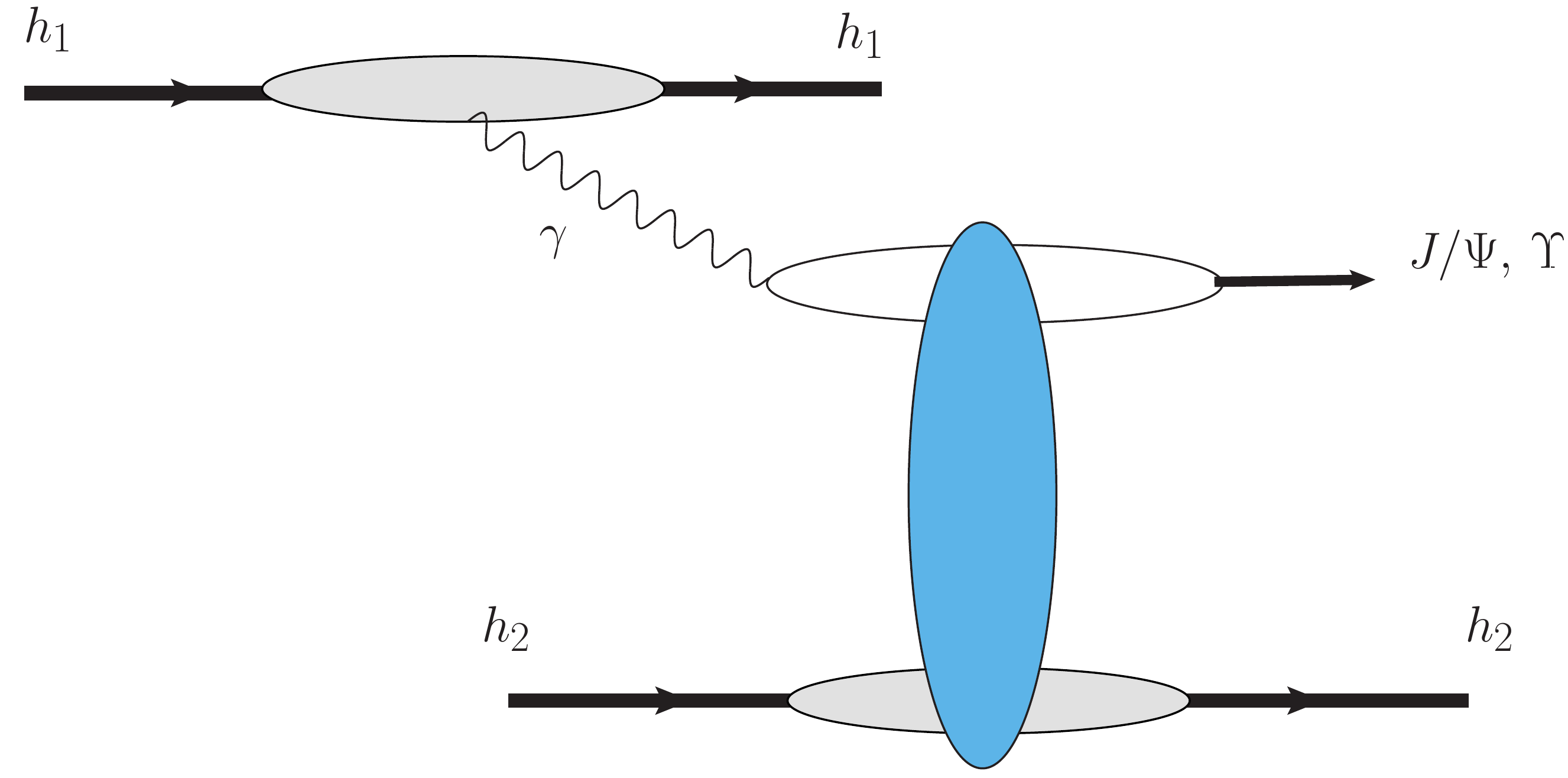} \\
	(a) & (b)
	\end{tabular}
	\caption{Exclusive vector meson photoproduction in (a) electron - hadron and (b) ultraperipheral collisions.   }
	\label{fig:diagram}
\end{figure}

One of the goals of particle physics is to achieve a deeper knowledge of the hadronic structure. An important phenomenological and experimental tool for this purpose is the deep inelastic $ep$ scattering (DIS), where an electron emits a virtual photon which interacts with a proton target, probing its partonic structure. A detailed experimental study of DIS was carried out at HERA, where the $\gamma p$ c.m. energy ($W$) reached a maximum value of the order of 200 GeV, and the data have shown that the gluon density inside the proton grows with the energy and that a nonnegligible fraction of  events ($\approx 10 \%$) is characterized by an intact proton in the final state (For a review see, e.g. Ref. \cite{Newman:2013ada}). In particular, HERA has measured the exclusive $J/\Psi$ and $\Upsilon$ production, represented in Fig. \ref{fig:diagram} (a), and observed that the associated cross sections have a steep power - like increasing with the energy, in agreement with the theoretical expectation that the exclusive vector meson  cross section is proportional to the square of the gluon distribution \cite{Ryskin:1992ui,Brodsky:1994kf}. Such strong dependence on the underlying QCD dynamics has motivated an intense phenomenology over the last decades, with the search for the nonlinear QCD effects \cite{hdqcd} being one of the major motivations for the construction of the Electron - Ion Collider (EIC) in the USA \cite{eic}, recently approved, as well as for the proposal of future electron -- hadron colliders at CERN \cite{lhec}. These colliders are expected to allow the investigation of the hadronic structure at high energies with unprecedented precision to inclusive and diffractive observables. An alternative, originally proposed in Refs. 
\cite{Klein:1999qj,Goncalves:2001vs}, is to study the exclusive vector meson photoproduction in ultraperipheral collisions (UPCs), as represented in Fig. \ref{fig:diagram} (b).  In these collisions, two charged hadrons (or nuclei) interact at impact parameters larger than the sum of their radii \cite{upc}. Under these circunstances, it is well known that the hadron acts as a source of almost real photons and photon-hadron interactions may happen. As the maximum photon -- hadron center - of - mass energies reached in $pp$, $pPb$ and $PbPb$ collisions at the LHC and in the Future Circular Collider (FCC) \cite{fcc} are larger  than  those achieved at HERA,
the study of  different final states in UPCs allow us to improve our understanding of the QCD dynamics in an unexplored high energy  regime. 

In recent years, several theoretical works related to the exclusive vector meson photoproduction have been published and a great amount of data from HERA, RHIC and LHC has been accumulated. Currently, there is the expectation that the analysis of this process  in the next run of the LHC and in the future colliders  EIC, LHeC and FCC, will allow us to constrain the description of the exclusive $J/\Psi$ and $\Upsilon$ photoproduction in the high energy regime (For recent studies see, e.g. Refs. \cite{Goncalves:2020vdp,Goncalves:2020ywm}). In contrast, the amount of experimental data for low energies is scarce, in particular for energies close to the threshold and  $\Upsilon$ production, and the description of the process in this energy range is still theme of intense debate 
\cite{Kharzeev:1998bz,Brodsky:2000zc,Redlich:2000cb,Frankfurt:2002ka,Gryniuk:2016mpk,Hatta:2019lxo,Gryniuk:2020mlh,Xu:2020uaa,Zeng:2020coc,Du:2020bqj,Hatta:2019ocp,Boussarie:2020vmu,Mamo:2019mka}. A precise determination of the $J/\Psi$ and $\Upsilon$ photoproduction at low energies has been motivated by several aspects. Firstly, theoretical studies indicate that the near - threshold production of heavy quarkonium is sensitive to the trace anomaly contribution to the nucleon mass, which is one of the main open questions in hadronic physics \cite{Ji:1995sv,Hatta:2018sqd,Lorce:2017xzd,Hatta:2018ina,Wang:2019mza,Metz:2020vxd,Ji:2021pys,Ji:2021mtz,Kharzeev:2021qkd,Wang:2021dis,Kou:2021bez}. Second, it is expected to be possible to extract from the quarkonium photoproduction cross section near threshold the quarkonium - hadron scattering length, which can be used to determine the $J/\Psi$ and $\Upsilon$ binding energy in nuclear matter \cite{Gryniuk:2016mpk,Gryniuk:2020mlh,Du:2020bqj,Strakovsky:2019bev,Pentchev:2020kao}. Finally, this process is an irreducible background for the searching of the $P_c$ and $P_b$ pentaquark states, predicted to be produced  in photon - hadron interactions and to decay into the $J/\Psi \, p$ and $\Upsilon \, p$ final states, respectively \cite{vicmiguel2,Xie:2020niw,Xie:2020wfe,Xie:2020ckr,Cao:2019gqo}. Recently, the GlueX Collaboration at Jefferson Laboratory (JLab) has reported its data for the photoproduction of $J/\Psi$ in $ep$ scattering \cite{Ali:2019lzf}, with the new data for the threshold cross section being significantly different from the 40-year-old Cornell data and more data from JLab are expected in  a near future. For the $\Upsilon$ production, no data are available so far in the threshold region. Recent studies have demonstrated that the study of the exclusive $\Upsilon$ photoproduction is feasible in $ep$ collisions at the EIC \cite{Gryniuk:2020mlh} as well in the proposed electron ion collider in China (EicC) \cite{Xu:2020uaa}, which is expected to reach center - of - mass energies of the order of 20 GeV. In  Ref. \cite{Hatta:2019lxo}, the authors have demonstrated that the threshold production can be also studied in UPCs at RHIC in future, with the main challenge being to measure the quarkonia at very forward rapidities.

In this paper we will investigate the near threshold exclusive $J/\Psi$ and $\Upsilon$ photoproduction in $ep(A)$ collisions at the EicC and in ultraperipheral collisions at the LHC considering three different phenomenological models for the treatment of the vector meson production at low energies, which are based on distinct assumptions for the production close to the threshold. One of our goals is to estimate the current theoretical uncertainty and if future experimental data can be used to constrain the modeling of the process. Our predictions for the EicC will complement the studies performed in Refs. \cite{Xu:2020uaa,Zeng:2020coc} by presenting the results for the total cross sections, rapidity and energy distributions derived considering distinct phenomenological models. Another goal is to investigate, for the first time, the near threshold vector meson production in fixed - target collisions at the LHC \cite{after}. The studies performed in Refs. \cite{Lansberg:2015kha, Goncalves:2015hra,vicmiguel,Lansberg:2018fsy} have demonstrated that differently from the hadronic collisions at the LHC in the collider mode, which allow us to study the vector meson photoproduction at high energies, the fixed - target collisions allow the probe the photon - induced interactions in a  limited energy range, dominated by low - energy interactions, which is the region of interest for the study of the near threshold production. In Refs. \cite{vicmiguel2,Xie:2020ckr}, the photoproduction of pentaquark states in fixed target collisions at the LHC was investigated, and the results pointed out that a future experimental analysis of the $P_c$ and $P_b$ states is, in principle, feasible. However, in order to constrain the properties of these states, it is fundamental to have precise estimates of the exclusive vector meson photoproduction cross section at low energies, which is an irreducible background for the pentaquark  photoproduction. We will evaluate the rapidity distributions and total cross sections considering different projectile - target configurations assuming three distinct phenomenological models for the vector meson production. As we will demonstrate, the associated cross sections are large and the maxima of the rapidity distributions occur in the rapidity range probed by the LHCb detector. Moreover, our results indicate that the study of the exclusive vector meson photoproduction in fixed target collisions at the LHC can be useful to constrain the description of the near threshold production, as well as to investigate the origin of the proton mass.

The paper is organized as follows. 
In the next Section we present a brief review of the  formalism needed to describe the exclusive $J/\Psi$ and $\Upsilon$ photoproduction in $ep$ and ultraperipheral collisions, represented by the diagrams  shown in Fig. \ref{fig:diagram} (a) and (b), respectively. In particular, we will discuss the distinct phenomenological models considered for the description of the vector meson production at low energies, close to the threshold. In Section \ref{res}, we will present our predictions for the total cross sections and associated distributions considering $ep(A)$ collisions at the EicC and ultraperipheral collisions at the LHC in the fixed target mode. Finally, in Section \ref{conc} we will summarize our main results and conclusions.

\section{Formalism}
\label{form}
In order to describe the exclusive vector meson photoproduction in electron - hadron and ultraperipheral hadronic collisions, we will assume the validity of the Equivalent Photon Approximation (EPA) \cite{Budnev:1974de}, which allow us to factorize the  cross sections in terms of the photon fluxes, associated to the electron and to the hadron,  and the photon - hadron cross section. In particular, for the exclusive production of a vector meson $V$ in $eh$ collisions, with $h = p$ or $A$,  the cross section  can be written as
\cite{Lomnitz:2018juf,Klein:2019avl}
\begin{equation}
\sigma (eh\rightarrow eVh)= \int {dk} \int dQ^{2} \, \frac{d^{2}N}{dkdQ^{2}} \,
\sigma _{\gamma ^{\ast }h\rightarrow Vh}(W_{\gamma h},Q^{2}),
\end{equation}
where $k$  and $Q^2$ are the photon energy and virtuality, $W_{\gamma h}$ is the photon - hadron center of mass energy, $d^{2}N/dkdQ^{2}$ is the photon flux and $\sigma _{\gamma ^{\ast }h\rightarrow Vh}(W_{\gamma h},Q^{2})$ is the $\gamma h$ cross section.
One has  that the photon flux associated to the electron in the target rest frame is given by \cite{Budnev:1974de}
\begin{equation}
\frac{d^{2}N}{dkdQ^{2}}=\frac{\alpha }{\pi kQ^{2}}\Big[1-\frac{k}{%
	E_{e}}+\frac{k^{2}}{2E_{e}^{2}}-\Big(1-\frac{k}{E_{e}}\Big)\Big|\frac{%
	Q_{min}^{2}}{Q^{2}}\Big|\Big]\,\,,
\end{equation}
where $\alpha$ is the electromagnetic fine structure constant, $E_e$ is the energy of the initial electron and  $Q^2_{min} = m_e^2k^2/(E_e(E_e-k))$. It is important to emphasize that the main contribution for the photon flux comes from events with small values of $k$ and $Q^2$. As in Refs. \cite{Lomnitz:2018juf,Klein:2019avl}, we will assume that the $Q^{2}$ dependence of the $\gamma h$ cross section can be factorized as follows
\begin{equation}
\sigma _{\gamma ^{\ast }h\rightarrow Vh}(W_{\gamma h},Q^{2})=\sigma _{\gamma
	h\rightarrow Vh}(W_{\gamma h},Q^{2}=0)\bigg(\frac{M_{V}^{2}}{M_{V}^{2}+Q^{2}}\bigg)%
^{\eta },
\end{equation}
where $M_V$ is the mass of the vector meson and $\eta = c_1+c_2(Q^2+M_V^2)$, with  the values of $c_1$ and $c_2$  being determined by fitting the HERA data for the $J/\Psi$ production. In our analysis we will assume that $c_1 = 2.36$ and $c_2$ = 0.0029 $\mathrm{GeV}^{-2}$ \cite{Lomnitz:2018juf} for both mesons. As we will focus on events with small virtualities ($Q^2 \le 1.0$ GeV$^2$), the impact of the above assumption on our predictions is negligible.

For ultraperipheral hadron - hadron collisions, the incident hadrons can be considered as sources of photons, which are assumed to be coherently radiated by the whole hadron. Such condition imposes that the minimum photon wavelength must be greater than the hadron radius $R$ and, consequently, the photon virtuality must satisfy $Q^2 = -q^2 \le 1/R^2$. Therefore, in UPCs, the photon virtuality can be neglected and the photons can be considered as being real. Consequently, the exclusive  vector meson production cross section in UPCs can be written as
\begin{equation}
\sigma(h_1 h_2 \rightarrow h_1 V  h_2;\sqrt{s_{NN}}) =  \int dk \, \left.\frac{dN}{dk}\right|_{h_1} 
\cdot \sigma_{\gamma h_2 \rightarrow V \, h_2}\left(W_{\gamma h_2}  \right) + \int dk \, \left.\frac{dN}{dk}\right|_{h_2} \cdot  \sigma_{\gamma h_1 \rightarrow V \, h_1}\left(W_{\gamma h_1}  \right)\,  \; , 
\label{eq:sigma_pp}
\end{equation}
where $\sqrt{s_{NN}}$ is the nucleon - nucleon center - of - mass energy and $W_{\gamma h} = [2 \,k \, \sqrt{s_{NN}}]^{1/2}$. Moreover,  the photon flux associated to the hadron $h$ is given by:
\begin{eqnarray}
\left.\frac{dN}{dk}\right|_{h} = \int \mbox{d}^{2} {\mathbf b} \, P_{NH} ({\mathbf b}) \,   N\left(k,{\mathbf b}\right) \,\,, 
\end{eqnarray}
where $P_{NH} ({\mathbf b})$ is the probability of not having a hadronic interaction at impact parameter ${\mathbf b}$  and the number of photons per unit area, per unit energy, derived assuming a point-like form factor, is given by 
\begin{equation}
N(k,{\mathbf b}) = \frac{Z^{2}\alpha}{\pi^2} \frac{k}{\gamma^{2}}
\left[K_1^2\,({\zeta}) + \frac{1}{\gamma^{2}} \, K_0^2({\zeta}) \right],\,
\label{fluxo}
\end{equation}
where $\gamma$ is the Lorentz factor, $\zeta \equiv k b/\gamma$ and $K_0(\zeta)$ and  $K_1(\zeta)$ are the
modified Bessel functions. As in Ref. \cite{Klein:2016yzr}, we will assume $P_{NH} ({\mathbf b}) = \exp[-\sigma_{NN} \, T_{AA} ({\mathbf b})]$ in nuclear collisions, where $\sigma_{NN}$ is the nucleon - nucleon interaction cross section and $T_{AA} ({\mathbf b})$ is the nuclear overlap function. For proton - nucleus collisions, $P_{NH} ({\mathbf b}) = \exp[-\sigma_{NN} \, T_{A} ({\mathbf b})]$, where $T_{A} ({\mathbf b})$ is the nuclear thickness function (For details see Ref. \cite{Klein:2016yzr}). 

The main ingredient to estimate the  vector meson production in $ep$ and ultraperipheral hadronic collisions is the photoproduction cross section $\sigma _{\gamma h\rightarrow Vh}$. In  our analysis, we will consider the distinct treatments for the heavy vector meson production at low energies ($W \le  20$ GeV), with particular focus on $\gamma h$ center - of - mass energies close to the threshold. Twenty years ago, in Ref. \cite{Brodsky:2000zc}, the authors have proposed a model for the quarkonium production in the threshold regime, motivated to the possibility of use this process to probe multiquark, gluonic and hidden - color correlations in the hadronic wavefunction in QCD. 
In this model, the near threshold exclusive $\gamma p\rightarrow Vp$ production  is described in terms of the two - gluon exchange, and the differential cross section takes the form \cite{Brodsky:2000zc}
\begin{eqnarray}
\left.\frac{d\sigma}{dt}\right|_{\gamma h\rightarrow Vh} = N_{V}v\frac{(1-x)^2}{R^2M_V^2}F^2(t) (W_{\gamma p}^2-m_p^2)^2,
\label{Eq:twogluon}
\end{eqnarray}
where $x=(2m_pM_V+M_V^2)/(W_{\gamma p}^2-m_p^2)$, $R = 1.0$ fm and $v=1/16\pi (W_{\gamma p}^2-m_p^2)^2$.
$F(t)$ is the proton form factor that takes into account of the recombination of the outgoings quarks into the final proton after the gluon emission. Following Ref. \cite{Brodsky:2000zc},  we will assume $F^2(t) = \exp(bt)$, with  $b=1.13 $ ($b=1.67$) GeV$^{-2}$       for the $J/\Psi \,(\Upsilon)$ production. Moreover, $N_V$ is the normalization factor, which is adjusted to the experimental data at low energies. The predictions associated to this model will be denoted Two - gluon hereafter. It is important to emphasize that this model is expected to be valid in the threshold regime, since at larger energies, higher order corrections associated to the QCD dynamics become important and should be considered. In our analysis, we will also estimate the photoproduction cross section on the proton considering a model based on the Vector Meson Dominance (VMD) \cite{sakurai}. Over the years, several authors have considered such approach, mainly motivated by the possibility of improve our understanding of the quarkonium - proton interaction, which is needed to describe the quarkonium suppression in heavy ion collisions as well as to estimate the associated scattering length (See, e.g. Refs. \cite{Kharzeev:1998bz,Redlich:2000cb,Gryniuk:2016mpk,Gryniuk:2020mlh}). In the VMD approach, the  $\gamma p\rightarrow Vp$ scattering can be described in terms of the elastic $V p \rightarrow V p$  cross section, denoted $\sigma^{el}_{Vp}$, as follows
\begin{eqnarray}
\sigma_{\gamma h\rightarrow Vh} = \bigg(\frac{ef_{V}}{M_V}\bigg)^2 \bigg(\frac{q_{Vp}}{q_{\gamma p}}\bigg)^2\,\sigma^{el}_{Vp}\,,
\end{eqnarray}
where $f_V$ is the vector meson decay constant and $q_{Vp}$ ($q_{\gamma p}$) denotes the magnitude of the vector meson (photon) three momentum in the c.m. frame of the $V p \rightarrow V p$ ($\gamma p\rightarrow Vp$) process. Recently, the VMD model was applied in Refs. \cite{Gryniuk:2016mpk,Gryniuk:2020mlh} to estimate the quarkonium - proton scattering length from $\gamma p \rightarrow V p$ experiments, with particular emphasis on the future EIC. The final expression for the $\gamma p\rightarrow Vp$ cross section is given by
\begin{eqnarray}
\sigma_{\gamma p\rightarrow Vp} = \bigg(\frac{ef_{V}}{M_V}\bigg)^2\frac{C^V_{el}}{2 W_{\gamma p} q_{\gamma p}}\bigg(\frac{q_{Vp}}{q_{\gamma p}}\bigg)
\bigg(1-\frac{\nu_{el}}{\nu}\bigg)^{b^V_{el}}\bigg(\frac{\nu}{\nu_{el}}\bigg)^{a^V_{el}},
\end{eqnarray}
where  $\nu$ is defined as 
\begin{eqnarray}
\nu = \frac{1}{2}(W_{\gamma p}^2-m_p^2-M_V^2)
\end{eqnarray}
and $\nu_{el}=m_p M_V$ corresponds to the elastic threshold. In addition, the parameters $a^V_{el}$, $b^V_{el}$ and $C^V_{el}$ are determined by fitting the available data points for the exclusive vector meson production cross sections (For details see  Refs. \cite{Gryniuk:2016mpk,Gryniuk:2020mlh}). In what follows, we will denote the predictions derived using this model by VMD. Finally, the exclusive vector meson photoproduction can also be estimated considering a phenomenological model inspired in Regge theory, with the cross section being usually parameterized by 
\begin{eqnarray}
\sigma_{\gamma p \rightarrow V  p} = \sigma_{\pom} \times W_{\gamma p}^{\epsilon} + \sigma_{\reg} \times W_{\gamma p}^{\eta} \,\,,
\label{sig_gamp}
\end{eqnarray}
where the first term is associated to a Pomeron exchange and the second one to the Reggeon exchange. Such parameterization is implemented in the STARlight \cite{Klein:2016yzr} and eSTARlight \cite{Lomnitz:2018juf}  event generators. In contrast to the light meson production, where both terms are assumed to contribute, the $J/\Psi$ and $\Upsilon$ productions are modeled by assuming that the interaction is described only by the Pomeron contribution, with the cross section being supplemented by a factor that accounts for its behavior for energies near  the  threshold of production. As a consequence, the heavy vector meson photoproduction cross section is described by 
\begin{eqnarray}
\sigma_{\gamma p \rightarrow V  p} = \sigma^V_{\pom} \times W_{\gamma p}^{\epsilon_V} \times \left(1-\frac{(M_V+m_p)^2}{W_{\gamma p}^2}\right)  \,\,,
\end{eqnarray}
with the free parameters on the parameterization, $\sigma^V_{\pom}$ and $\epsilon_V$, being fitted using the HERA data for the exclusive $J/\Psi$ and $\Upsilon$ production. Such model will be denoted by Regge model hereafter.

In our analysis we also will consider the exclusive vector meson photoproduction in photon - nucleus interactions at the EicC and LHC. For the $J/\Psi$ production, we will derive the $\gamma A \rightarrow J/\Psi A$ cross section using the optical theorem, the vector dominance model  \cite{sakurai} and the classical Glauber approach \cite{glauber}, which implies that  the forward differential cross section for a photon - nucleus interaction is given by
\begin{eqnarray}
\frac{d\sigma(\gamma A \rightarrow J/\Psi A)}{dt}|_{t=0} = \frac{\alpha \sigma^2_{tot}(J/\Psi A)}{4 f_{\psi}^2} \,\,,
\label{gvdm}
\end{eqnarray}
where $f_{\psi}$ is the $J/\Psi$ -- photon coupling and the total cross section for the $J/\Psi$ -- nucleus interactions is expressed as follows
\begin{eqnarray}
\sigma_{tot}(J/\Psi A) = \int \mbox{d}^{2} {\mathbf b} \,\,\{ 1 - \exp[-\sigma_{tot}(J/\Psi p)T_{AA}({\mathbf b})]\},
\label{glauber}
\end{eqnarray}
with  $T_{AA}$ being the overlap function at a given impact parameter ${\mathbf b}$ and $\sigma_{tot}(J/\Psi p)$ is determined by  $\sigma(\gamma p \rightarrow J/\Psi p)$ (See Eq. (9) in Ref. \cite{Klein:1999qj}). The total cross section will be given by
\begin{eqnarray}
\sigma_{\gamma A \rightarrow J/\Psi A} = \int_{t_{min}}^{\infty} dt \, \frac{d\sigma(\gamma A \rightarrow J/\Psi A)}{dt}|_{t=0} \,|F_A(t)|^2 \,\,,
\label{sig_gamA}
\end{eqnarray}
where $F_A(t)$ is the nuclear form factor and $t_{min} = (M_{\psi}^2/4 k \gamma)^2$. On the other hand, for the $\Upsilon$ production we will assume the impulse approximation, which implies  $\sigma(\gamma A\to \Upsilon A)=A\cdot \sigma(\gamma p\to \Upsilon p)$. Such assumption is a reasonable first approximation since the $\Upsilon$ state is a  compact object, with negligible interactions with the nuclear medium. In the next Section we will estimate the heavy vector meson photoproduction in $\gamma p$ and $\gamma A$ interactions considering the predictions of the Two - gluon, VMD and Regge models for $\sigma_{\gamma p \rightarrow V  p}$ as input in the calculations.

\begin{figure}[t]
	\centering
	\includegraphics[width=0.45\textwidth]{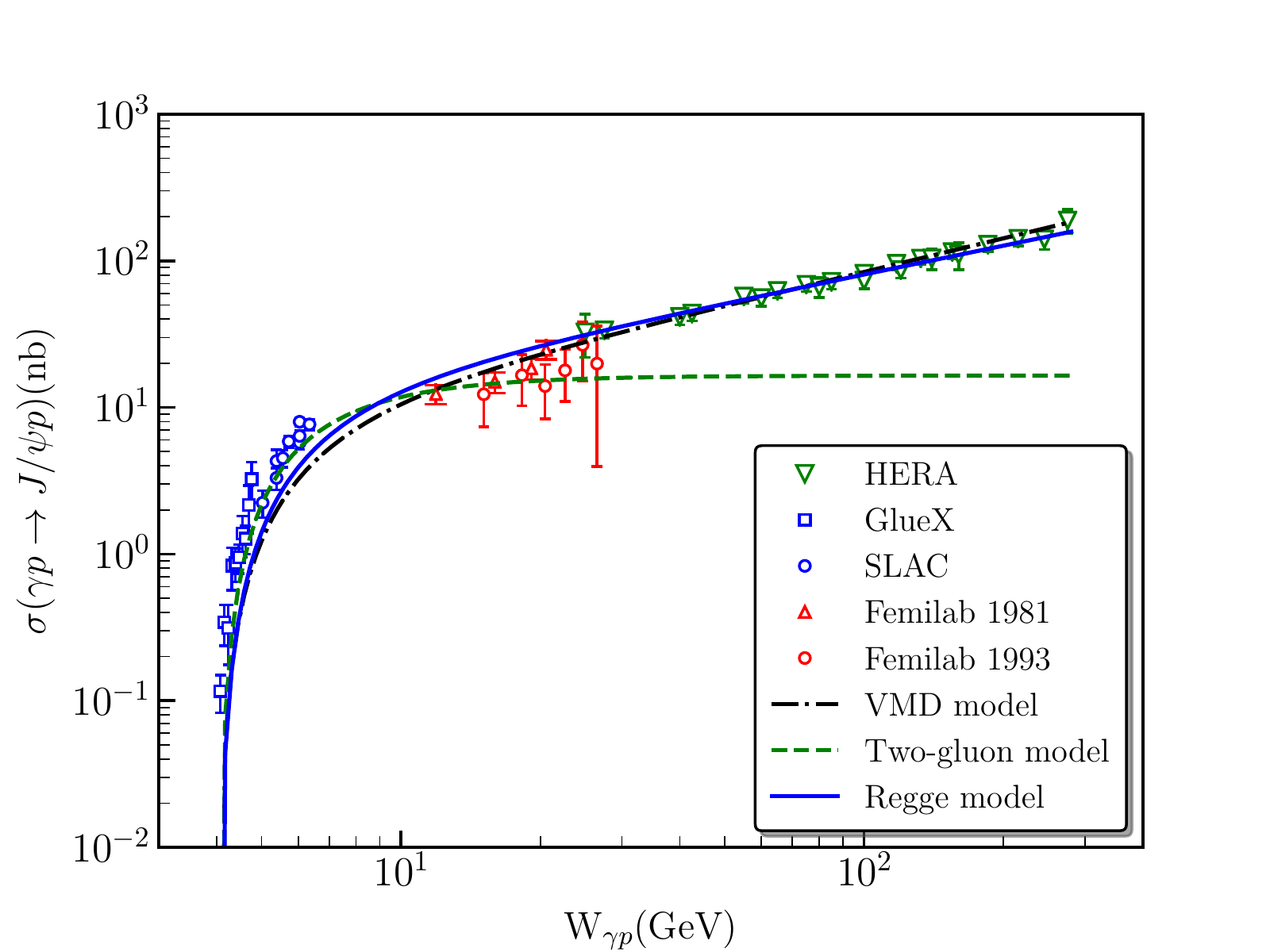}
	\includegraphics[width=0.45\textwidth]{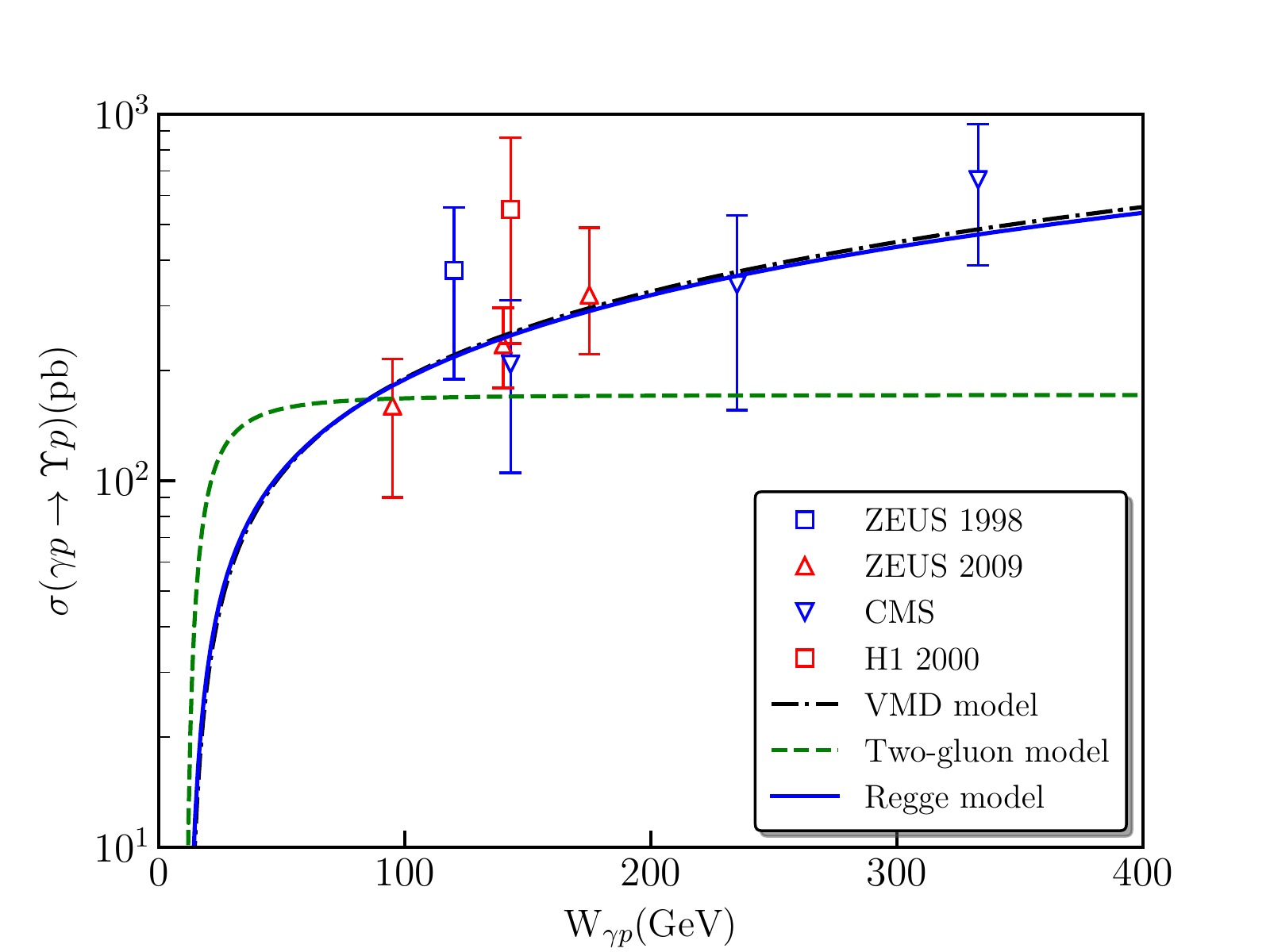}
	\caption{Energy dependence of the $\gamma p \rightarrow J/\Psi p$ (left panel) and $\gamma p \rightarrow \Upsilon p$ (right panel) total cross sections predicted by the Two - gluon, VMD and Regge models. Data from Refs.  \cite{Chekanov:2002xi,Ali:2019lzf,Camerini:1975cy,Binkley:1981kv,Frabetti:1993ux,Breitweg:1998ki,Adloff:2000vm,Chekanov:2009zz,Sirunyan:2018sav}. }
	\label{Fig:data}
\end{figure}

\section{Results}
\label{res}
In this Section we will present our predictions for the exclusive $J/\Psi$ and $\Upsilon$ photoproduction in fixed - target collisions at the LHC and  $ep$ and $eAu$ collisions at the EicC. The main input in our calculations is the ${\gamma p \rightarrow V  p}$ cross section, which will be described by the Two - gluon, VMD and Regge models discussed in the previous Section. In our study we will employ the STARlight \cite{Klein:2016yzr} and eSTARlight \cite{Lomnitz:2018juf} event generators, which were modified by the inclusion of the Two - gluon and VMD models. In Fig.  \ref{Fig:data} we present the energy dependence of the $J/\Psi$ (left panel) and  $\Upsilon$ (right panel) cross sections predicted by these distinct models. The current  experimental data is also presented for comparison.
One has that the VMD and Regge predictions are similar, which can be explained by the fact that in both models the energy dependence of the cross sections is power - like, with  the free parameters being adjusted by the same set of data. On the other hand, the Two - gluon model provides a satisfactory description of the  near threshold $J/\Psi$ photoproduction but fails to describe the data for $W \gtrsim 20 $ GeV. As discussed in the previous Section, such result is expected since the QCD evolution is not taken into account in this model. Our results for $J/\Psi$ indicate  that the  predictions of the three models  differ for $W \lesssim 20 $ GeV and, consequently, new data in this region is fundamental to improve the description of the near threshold production. For the $\Upsilon$ production, the behaviour of the cross section for the near threshold region is still an open question, since no data are available so far. As a consequence, it is not clear the energy range of validity of the Two - gluon model. Following Ref. \cite{Cao:2019gqo}, the normalization $N_{\Upsilon}$ in Eq. (\ref{Eq:twogluon}) will be adjusted using the data for $W_{\gamma p}$ around 100 GeV, which is the data for the lowest center - of - mass energy. That implies that the Two - gluon predictions for $\Upsilon$ production should be considered an upper bound.  Surely, new data for low energies is needed in order to constrain the near  threshold $\Upsilon$ photoproduction. Our goal, in what follows, is to verify if future experimental analyzes at LHC and EicC can probe this kinematical range and improve our understanding of the near threshold production.
\begin{center}
	\begin{table}[t]
		\begin{tabular}{|c|c|c|c|c|c|c|c|}
			\hline 
{\bf $\sigma_{h_1 h_2 \rightarrow h_1 J/\Psi h_2}$ [nb]	} & {\bf p - Pb} &  {\bf p - Ar} & {\bf Pb - p}  & {\bf Pb - He} & {\bf Pb - Ar} \tabularnewline
			\hline 
			\hline
			Regge model & 7.9$\times 10 ^2$ &  6.4 $\times 10 ^1$&  2.5$\times 10 ^2$   & 4.2$\times 10 ^2$  & 3.4$\times 10 ^3$   \tabularnewline
			\hline 
			VMD model & 6.6$\times 10 ^2$ &  5.4$\times 10 ^1$& 2.1$\times 10 ^2$  & 3.5$\times 10 ^2$    & 2.9$\times 10 ^3$   \tabularnewline
			\hline 
			Two-gluon model & 7.7$\times 10 ^2$  & 5.8 $\times 10 ^1$ & 2.7$\times 10 ^2$   & 3.6$\times 10 ^2$  & 2.8$\times 10 ^3$  \tabularnewline
			\hline 
			\hline 
{\bf $\sigma_{h_1 h_2 \rightarrow h_1 \Upsilon h_2}$ [pb]}	& &		&   &  &   \tabularnewline
			\hline 
	Regge model & 1.7$\times 10 ^2$ & 2.2 $\times 10 ^1$    & 1.3$\times 10 ^1$   & 1.4$\times 10 ^2$  & 8.7$\times 10^2$  \tabularnewline
			\hline 
			VMD model& 1.5$\times 10 ^2$ & 2.0$\times 10 ^1$ & 1.1$\times 10 ^1$  & 1.2$\times 10 ^2$    & 7.7$\times 10^2$   \tabularnewline
			\hline 
			Two-gluon model & 5.6$\times10 ^2$ & 6.9$\times 10 ^1$ & 4.8$\times 10 ^1$  & 4.5$\times 10 ^2$ & 31.0$\times 10^2$   \tabularnewline
			\hline 
			\hline
		\end{tabular}
		\caption{Total cross sections  for the exclusive $J/\Psi$ and $\Upsilon$ photoproduction in fixed - target collisions at the LHC considering different models for the $\gamma p \rightarrow V p$ cross section. We assume  $\sqrt{s_{NN}} = 100$ (69) GeV for $pA$ ($PbA$) collisions.  }
		\label{table:LHC}
	\end{table}
\end{center}
\begin{figure}[t]
	\centering
	\includegraphics[width=0.45\textwidth]{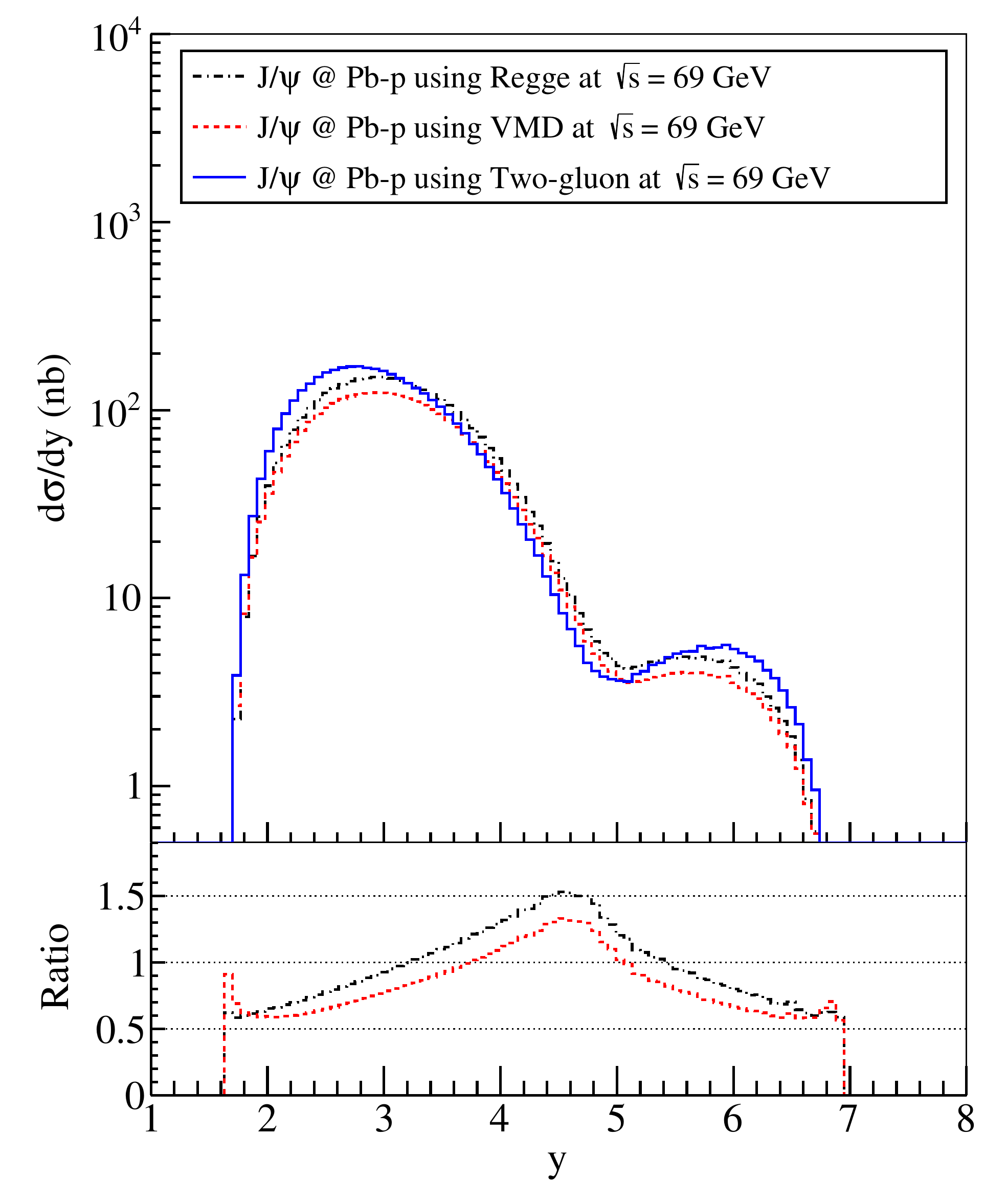}
	\includegraphics[width=0.45\textwidth]{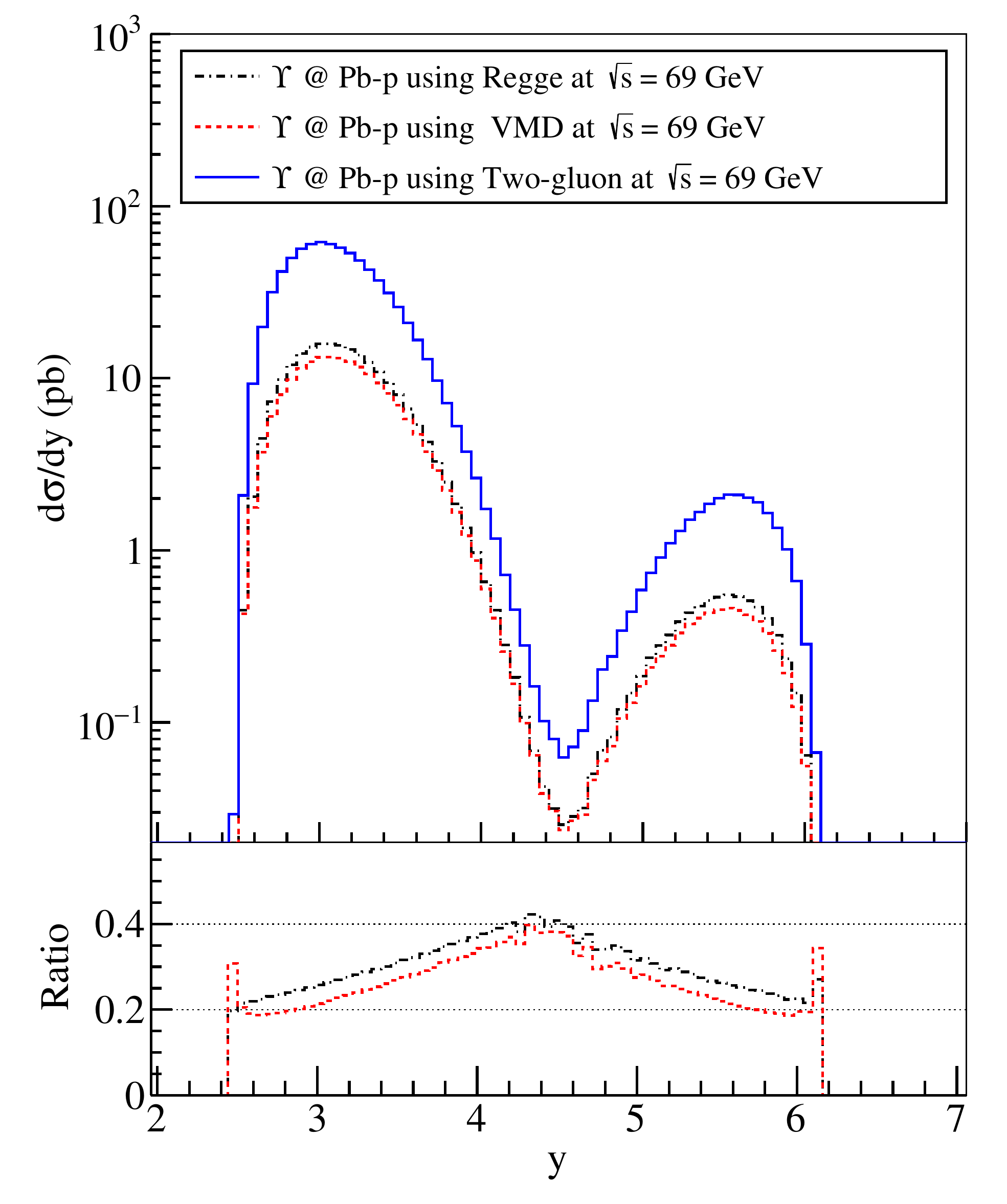}
	\caption{Rapidity distributions for the exclusive $J/\Psi$ (left panel) and  $\Upsilon$ (right panel) photoproduction in $Pbp$ collisions at  $\sqrt{s_{NN}} = 69$ GeV considering distinct models for the $\gamma p \rightarrow V p$ cross section.}
	\label{Fig:rapLHC}
\end{figure}

Initially, let's investigate the exclusive heavy vector meson photoproduction in fixed - target collisions at the LHC. During the last decade,  a large amount of data have been collected by the LHC considering $pp$, $pPb$ and $PbPb$ collisions in the collider mode for different center - of - mass energies \cite{upc}. 
In recent years, the possibility of study a complementary kinematical range in fixed - target collisions at the LHC was proposed \cite{after} and the analysis of these collisions became a reality by the injection of noble gases ($He, \, Ne, \, Ar$) in the LHC beam pipe by the LHCb Collaboration \cite{lhcfixed} using the System for Measuring Overlap with Gas (SMOG) device \cite{smog}. For  the typical fixed - target $pA$ and $PbA$ configurations, the center - of - mass energies reached were  $\sqrt{s_{NN}} \approx 110$ GeV  and $\sqrt{s_{NN}} \approx 69$ GeV, respectively, with the associated data having been  used to improve our understanding of the nuclear effects present in $pA$ collisions \cite{Aaij:2018ogq,Maciula:2020cfy} and, in the particular case of $pHe$ collisions, to shed light on the antiproton production (See e.g. Ref. \cite{antiproton}). Moreover, in Ref. \cite{vicmiguel}, the authors have pointed out that the study of the exclusive vector meson photoproduction  in fixed - target collisions is feasible and that the main contribution for the total cross section comes from the near threshold region. In what follows, we will investigate in more detail this conclusion and derive predictions for the total cross sections and rapidity distributions considering the Two - gluon, VMD and Regge models as input in our calculations. 

Our predictions for the total cross sections are presented in Table \ref{table:LHC} for different projectile - target configurations and assuming $\sqrt{s_{NN}} = 100$ (69) GeV for $pA$ ($PbA$) collisions. It is important to emphasize that although a $Pb$ target is not possible with the SMOG system, we decided to present the associated predictions since it can envisioned with a solid target and a bent crystal, as discussed in Ref. \cite{after}.
One has that in the $J/\Psi$ case, the Regge model provides the upper bound. In contrast, for the $\Upsilon$ production, this bound is provided by the Two - gluon model, which is expected from Fig. \ref{Fig:data}.
We predict values of the order of hundreds of nb (pb) for the $J/\Psi$ ($\Upsilon$) production in $pPb$ collisions at $\sqrt{s_{NN}} = 110$ GeV.   For $Pbp$ collisions at $\sqrt{s_{NN}} = 69$ GeV, our predictions are smaller by a factor $\gtrsim 3$.
In Ref. \cite{after}, the authors have discussed the expected luminosities for fixed - target collisions at the LHC considering different technological options for the target solution. In general, the  instantaneous luminosity obtained  in $pA$ collisions is larger than in the $PbA$ case and larger center - of - mass  energies  are reached when a proton beam is considered. One has that in $pA$ collisions at $\sqrt{s_{NN}} \approx 110$ GeV, the integrated luminosity is expected to be larger than $100 \, nb^{-1}$ per year. Consequently, we predict that the number of events per year will be larger than $5.4 \times 10^3$ ($3$) for the $J/\Psi$ ($\Upsilon$) photoproduction in $p  Ar$ collisions at $\sqrt{s_{NN}} = 110$ GeV.
On the other hand, for $PbA$ collisions at $\sqrt{s_{NN}} \approx 69$ GeV, the integrated  luminosity is strongly dependent on the target solution. For a SMOG - like device,  it is of the order of $0.01 \, nb^{-1}$ per year, but can reach higher values, ${\cal{O}}(100) \, nb^{-1}$ per year, for a gas-jet target. Therefore, we predict that the number of events per year for the $J/\Psi$ ($\Upsilon$) photoproduction in $Pb  Ar$ collisions at $\sqrt{s_{NN}} = 69$ GeV can reach values of the order  $2.8 \times 10^4$ ($7.7 \times 10^1$) for the higher luminosities. For other projectile - target configurations, we predict smaller values for the cross sections, which implies a strong reduction on  the number of events associated to the $\Upsilon$ production. However, our results indicate that the number of $J/\Psi$ events is still large for these other configurations. In particular, our results for $Pbp$ collisions indicate that the number of events will be high enough to allow a detailed analysis of the transverse momentum distributions, which is needed  to  estimate the proton mass radius (See e.g. Refs. \cite{Kharzeev:2021qkd,Kou:2021bez}).  

 In Fig. \ref{Fig:rapLHC} we present our predictions for the rapidity distributions associated to the exclusive $J/\Psi$ (left panel) and  $\Upsilon$ (right panel) photoproduction in $Pbp$ collisions at  $\sqrt{s_{NN}} = 69$ GeV. The distributions are asymmetric, which is expected since we are considering the collision of non - identical hadrons, which are characterized by distinct  magnitudes of the associated photon fluxes. Moreover, the maximum of the distributions occurs for forward rapidities and  in the kinematical range probed by the LHCb detector.  In the lower panels of Fig. \ref{Fig:rapLHC} we present our results for the ratios  between the predictions derived using the VMD and Regge models and that obtained using the Two - gluon model. One has that, for $y = 2.5$, the Two - gluon predictions for the $J/\Psi$ ($\Upsilon$) production  are a factor $\approx$ 2 (5) larger than those obtained using the VMD and Regge models. Moreover, the shape of the distributions depends on the model considered. Another important aspect is that the threshold for the $\Upsilon$ production occurs at $y = 2.4$, i.e. within the kinematical range probed by the LHCb detector. For the $J/\Psi$ production, it occurs at $y = 1.7$, which is not accessed in this detector. However,  future data for $y \ge 2$ will be very useful to constrain the description of the $\gamma p \rightarrow J/\Psi p$ cross section in the energy range $W_{\gamma p} \ge 5.0$ GeV, where the theoretical uncertainty is large and not constrained by the existing data (See Fig. \ref{Fig:data}).  All these results point out that a future experimental analysis of the exclusive vector meson photoproduction in fixed - target collisions at the LHC is, in principle, feasible and that the study of this process can be useful to constrain the modeling of the near threshold regime.

\begin{center}
	\begin{table}[t]
		\begin{tabular}{|c|c|c|c|c|c|c|}
			\hline 
			& \multicolumn{2}{c|}{$ J/\Psi$}          &        \multicolumn{2}{c|}{$ \Upsilon$}   \\
			\hline 	
			& $\sigma(ep \rightarrow e J/\Psi p)$  & $\sigma(eAu \rightarrow e J/\Psi Au)$ &$\sigma(ep \rightarrow e \Upsilon p)$ & $\sigma(eAu \rightarrow e \Upsilon Au)$\tabularnewline
			\hline 
			Regge model & 6.5$\times 10 ^{2}$ (3.2$\times 10 ^{7}$) & 8.3$\times 10 ^4$ (4.1$\times 10 ^{9}$) & 1.2$\times 10 ^{-1}$ (6.0$\times 10 ^{3}$)  & 2.5 $\times 10 ^1$ (1.2$\times 10 ^{6}$)\tabularnewline
			\hline 
			VMD model& 5.4$\times 10 ^{2}$ (2.5$\times 10 ^{7}$) & 7.1$\times 10 ^4$ (3.5$\times 10 ^{9}$)   & 1.0$\times 10 ^{-1}$ (5.0$\times 10 ^{3}$)  & 2.1 $\times 10 ^1$  (1.0$\times 10 ^{6}$) \tabularnewline
			\hline 
			Two-gluon model & 6.2 $\times 10 ^2$ (3.1$\times 10 ^{7}$) & 6.4$\times 10 ^4$ (3.1$\times 10 ^{9}$) & 4.4$\times10 ^{-1}$ (22.0$\times 10 ^{3}$)  & 9.2$\times 10 ^1$ (4.5$\times 10 ^{6}$) \tabularnewline
			\hline 
			\hline 
		\end{tabular}
		\caption{Total cross sections (in pb) and number of events per year (in parentheses) for the exclusive $J/\Psi$ and $\Upsilon$ photoproduction in $ep$ and $eAu$ collisions at the EicC derived assuming $\sqrt{s} = 16.7$ GeV and considering different models for the $\gamma p \rightarrow V p$ cross section. The integrated luminosity per year is assumed to be 50 fb$^{-1}$.}
		\label{table:EICSeC}
	\end{table}
\end{center}

\begin{figure}[t]
	\centering
	\includegraphics[width=0.45\textwidth]{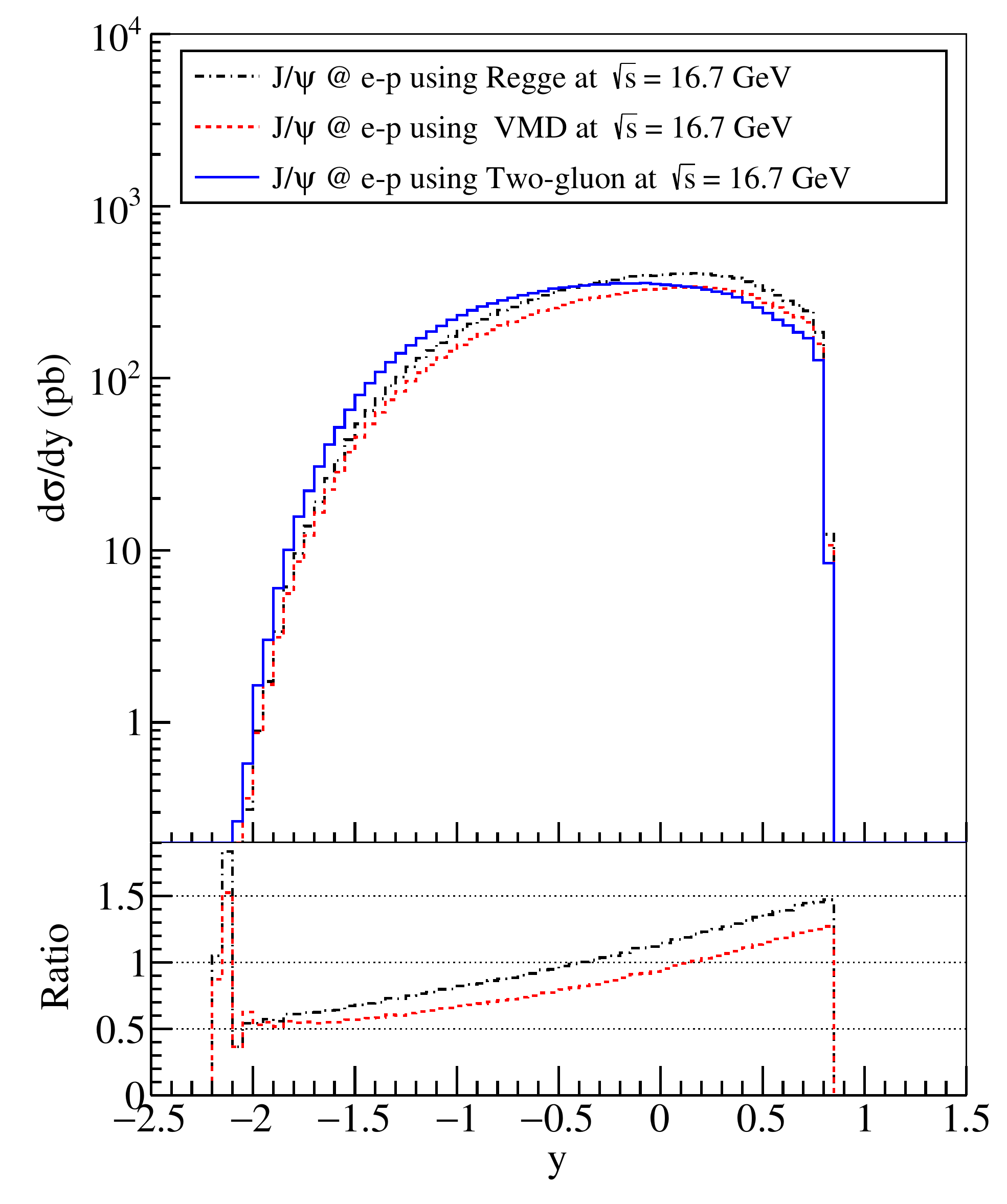}
	\includegraphics[width=0.45\textwidth]{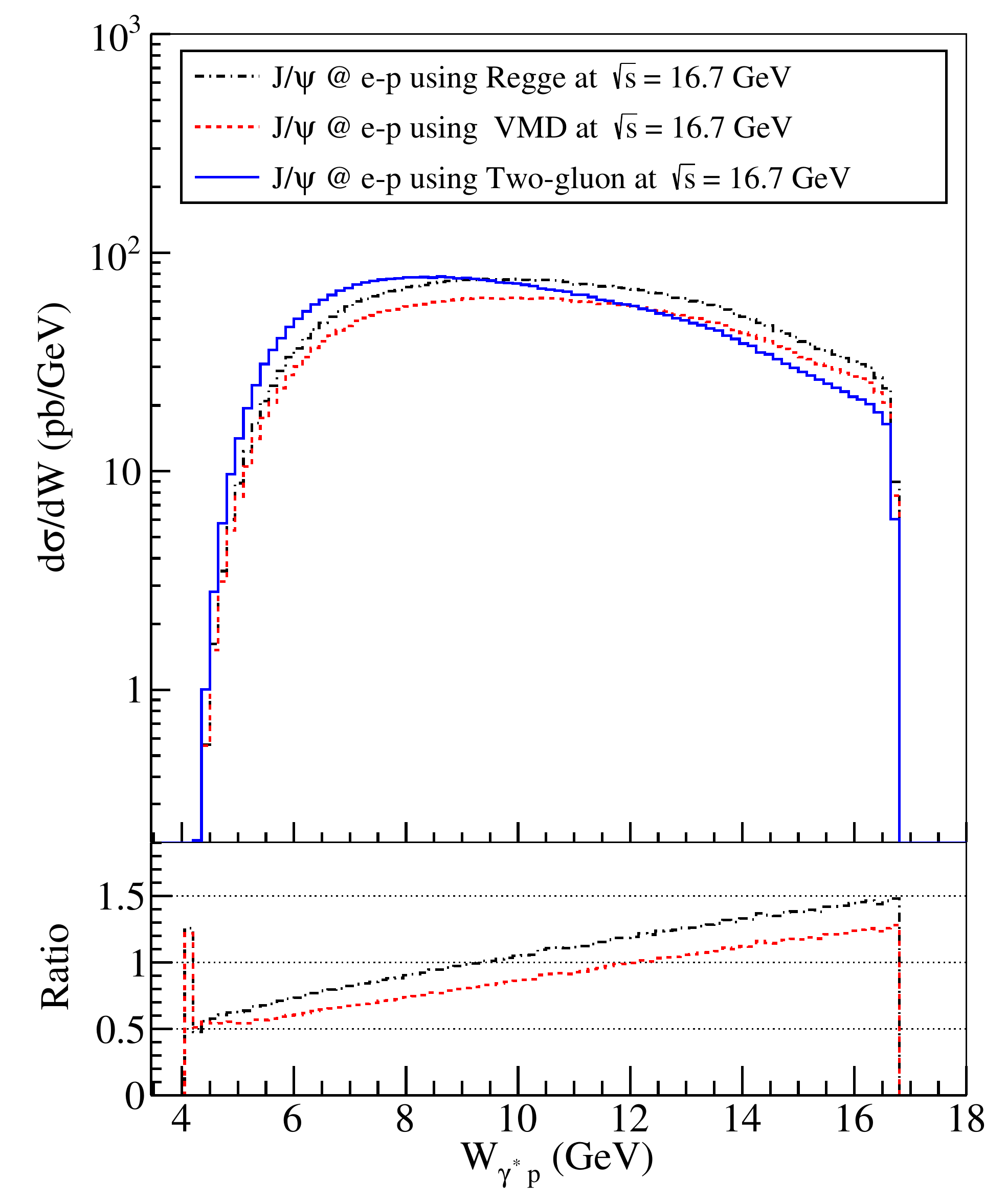} \\
		\includegraphics[width=0.45\textwidth]{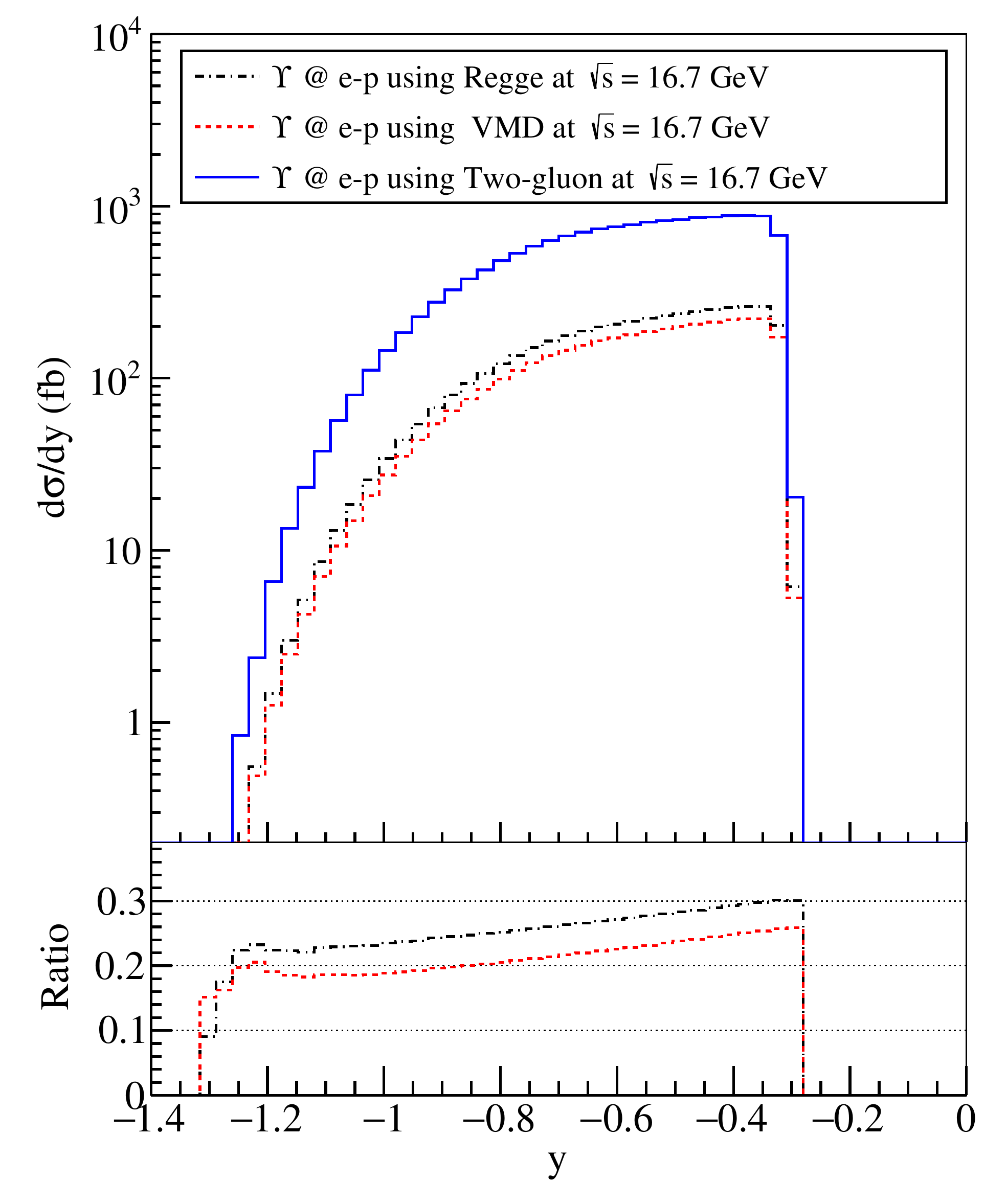}
	\includegraphics[width=0.45\textwidth]{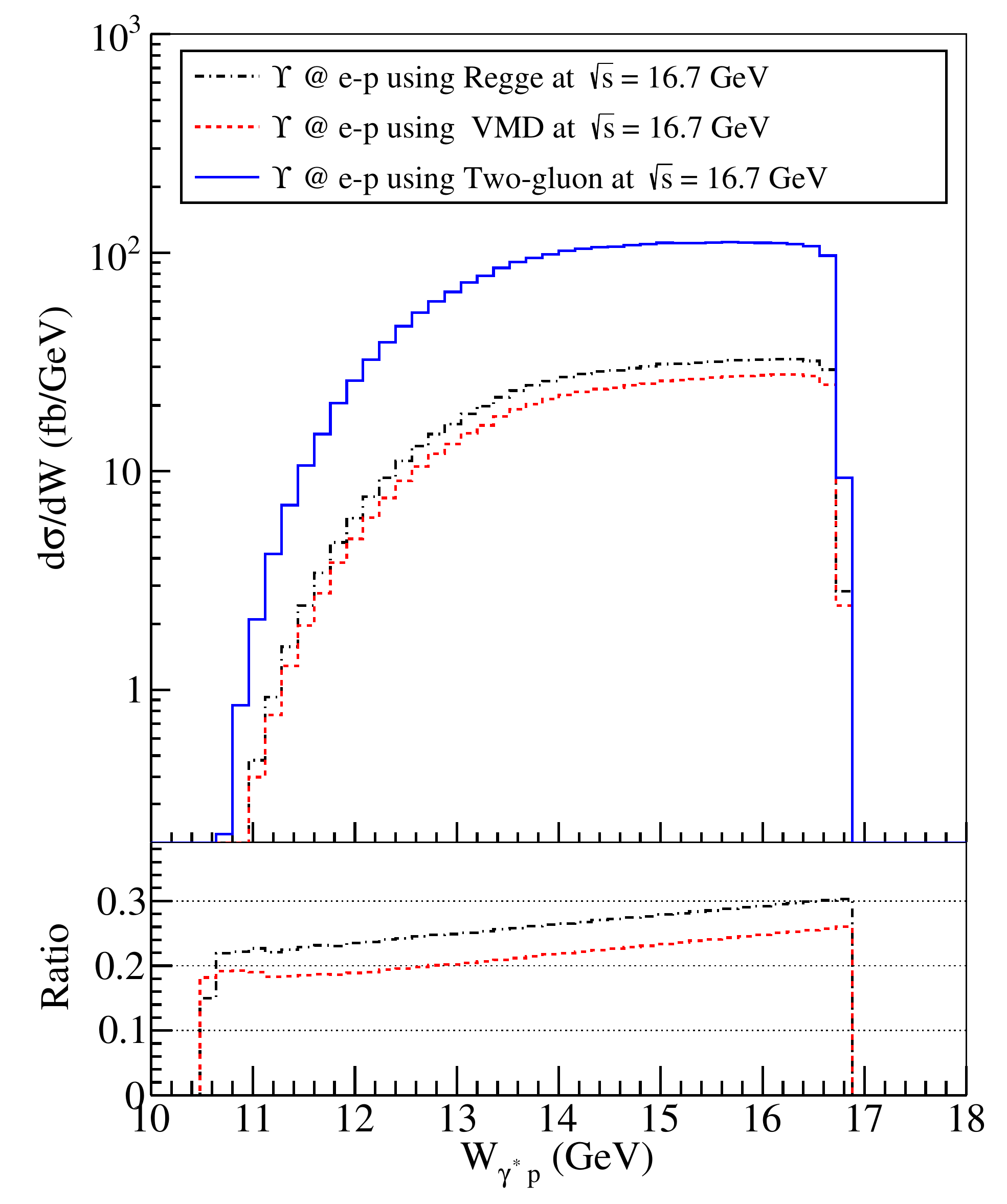}

	\caption{Predictions for the rapidity (left panels) and $W_{\gamma^* p}$ (right panels)  distributions associated to the exclusive $J/\Psi$ (upper panels) and $\Upsilon$ (lower panels) photoproduction in $ep$ collisions at the EicC ($\sqrt{s} = 16.7$ GeV).}
	\label{fig:EicC}
\end{figure}

 Furthermore, let's now estimate the exclusive vector meson photoproduction in $ep$ and $eAu$ collisions at the EicC. 
The proposed collider will provide highly polarized electrons and protons  with variable center of mass energies from
15 to 20 GeV and integrated luminosities per year of the order  of 50 fb$^{-1}$. Moreover,  unpolarized ion beams from Carbon to Uranium, will be also available at the EicC (For a detailed discussion see Ref. \cite{EicC}). Our results for the total cross sections and number of events per year in $ep$ and $eAu$ collisions at $\sqrt{s} = 16.7$ GeV, derived assuming $Q^2 \le 1$ GeV$^{2}$ and considering the Two-gluon, VMD and Regge models, are presented in Table \ref{table:EICSeC}. We predict a very large number of $J/\Psi$ events, which will allow to 
perform a detailed investigation of the near threshold production and, as a 
consequence, advance in our understanding about the QCD trace anomaly 
as well as in the searching of hidden - charm pentaquark states in the exclusive $J/\Psi$ photoproduction. 
In addition, our results for the $\Upsilon$ production indicate that the experimental analysis of this final state will also be feasible. As a consequence, a global data analysis of the $J/\Psi$ and $\Upsilon$ production  will be possible in the EicC, and it will allow to strongly reduce the theoretical uncertainties present in the treatment of the near threshold regime. Our predictions for the rapidity and $W_{\gamma p}$ distributions, derived considering $ep$ collisions at $\sqrt{s} = 16.7$ GeV, are presented in Fig. \ref{fig:EicC}. The results for the ratios between the VMD and Regge predictions and the Two-gluon one are also presented. One has the predictions of the distinct models differ on the normalization and that the shape of the distributions is also model dependent. Such result indicates that a future experimental analysis in the EicC can be useful to constrain the description of the vector meson photoproduction at low energies, in particular for energies close to the threshold.

\section{Summary}
\label{conc}
One of the main challenges of the strong interactions theory is the description of the origin of the nucleon mass. Recent results indicate that the nucleon mass receives a large contribution associated to the QCD trace anomaly and that it can be accessed through exclusive heavy vector meson photoproduction on a proton close to the threshold. In addition, this process provides important information about the quarkonium - proton cross section and is an irreducible background for the searching of pentaquark states in photon - hadron interactions. All these aspects strongly motivate a deeper understanding of the near threshold vector meson photoproduction. In this paper we have investigated the near threshold $J/\Psi$ and $\Upsilon$ photoproduction in fixed - target collisions at the LHC and in $ep (A)$ collisions at the EicC considering three distinct phenomenological models and demonstrated that future experimental analyzes in these colliders are feasible. In particular, fixed - target collisions at the LHC will allow us to perform a detailed investigation of the near threshold $J/\Psi$ photoproduction, which will be complementary to the current studies at the JLab.  For the EicC, 
we shown that  a detailed study of both final states can be performed, which will strongly diminish the theoretical uncertainty on the description of the near threshold production. Moreover, the EicC data for the $\Upsilon$ production will be complementary to that expected to be obtained in the EIC. Our results indicate that the study of the near threshold heavy vector photoproduction at the LHC and EicC is promissing and can be useful to estimate more precisely the contribution of the gluon condensate in the proton, which is closely related to QCD trace anomaly, as well as to improve the description  of the quarkonium - hadron scattering length, which is fundamental to describe the quarkonium propagation in a nuclear medium.

\section*{Acknowledgments}
The work is partially supported by the Strategic Priority Research Program of Chinese Academy of Sciences (Grant NO. XDB34030301). 
VPG was  partially financed by the Brazilian funding
agencies CNPq,   FAPERGS and  INCT-FNA (process number 
464898/2014-5).


\begin{thebibliography}{99}
\bibitem{Newman:2013ada}
P.~Newman and M.~Wing,
Rev. Mod. Phys. \textbf{86}, no.3, 1037 (2014)

\bibitem{Ryskin:1992ui}
M.~G.~Ryskin,
Z. Phys. C \textbf{57}, 89-92 (1993)

\bibitem{Brodsky:1994kf}
S.~J.~Brodsky, L.~Frankfurt, J.~F.~Gunion, A.~H.~Mueller and M.~Strikman,
Phys. Rev. D \textbf{50}, 3134-3144 (1994)





  \bibitem{hdqcd} 
  F.~Gelis, E.~Iancu, J.~Jalilian-Marian and R.~Venugopalan,
    Ann.\ Rev.\ Nucl.\ Part.\ Sci.\  {\bf 60}, 463 (2010);
  H.~Weigert,  Prog.\ Part.\ Nucl.\ Phys.\  {\bf 55}, 461 (2005); J.~Jalilian-Marian and Y.~V.~Kovchegov, Prog.\ Part.\ Nucl.\ Phys.\  {\bf 56}, 104 (2006)


\bibitem{eic}
  D.~Boer, M.~Diehl, R.~Milner, R.~Venugopalan, W.~Vogelsang, D.~Kaplan, H.~Montgomery and S.~Vigdor {\it et al.},
  arXiv:1108.1713 [nucl-th];
  A.~Accardi, J.~L.~Albacete, M.~Anselmino, N.~Armesto, E.~C.~Aschenauer, A.~Bacchetta, D.~Boer and W.~Brooks {\it et al.},
Eur.\ Phys.\ J.\ A {\bf 52}, no. 9, 268 (2016);  E.~C.~Aschenauer {\it et al.},
  Rept.\ Prog.\ Phys.\  {\bf 82}, no. 2, 024301 (2019); R.~Abdul Khalek, A.~Accardi, J.~Adam, D.~Adamiak, W.~Akers, M.~Albaladejo, A.~Al-bataineh, M.~G.~Alexeev, F.~Ameli and P.~Antonioli, \textit{et al.}
[arXiv:2103.05419 [physics.ins-det]].


\bibitem{lhec}
  J.~L.~Abelleira Fernandez {\it et al.}  [LHeC Study Group Collaboration],
  J.\ Phys.\ G {\bf 39}, 075001 (2012); P.~Agostini {\it et al.},
[arXiv:2007.14491 [hep-ex]].


\bibitem{Klein:1999qj} 
S.~Klein and J.~Nystrand,
Phys.\ Rev.\ C {\bf 60}, 014903 (1999)




\bibitem{Goncalves:2001vs}
V.~P.~Goncalves and C.~A.~Bertulani,
Phys. Rev. C \textbf{65}, 054905 (2002)



\bibitem{upc}
C. A. Bertulani and G. Baur, { Phys. Rep.} {\bf 163}, 299 (1988); F.~Krauss, M.~Greiner and G.~Soff,
  Prog.\ Part.\ Nucl.\ Phys.\  {\bf 39}, 503 (1997);
   C.~A. Bertulani, S.~R.~Klein and J.~Nystrand, Ann. Rev. Nucl. Part. Sci. {\bf 55}, 
271 (2005); V.~P.~Goncalves and M.~V.~T.~Machado,
  J.\ Phys.\ G {\bf 32}, 295 (2006);       A.~J.~Baltz {\it et al.},
  Phys.\ Rept.\  {\bf 458}, 1 (2008);       J.~G.~Contreras and J.~D.~Tapia Takaki,
  Int.\ J.\ Mod.\ Phys.\ A {\bf 30}, 1542012 (2015); 
      K.~Akiba {\it et al.} [LHC Forward Physics Working Group],
  J.\ Phys.\ G {\bf 43}, 110201 (2016); S.~R.~Klein and H.~Mantysaari,
Nature Rev. Phys. \textbf{1}, no.11, 662-674 (2019); S.~Klein and P.~Steinberg,
Ann.\ Rev.\ Nucl.\ Part.\ Sci.\  {\bf 70}, 323 (2020)



\bibitem{fcc} 
A.~Abada \textit{et al.} [FCC],
Eur. Phys. J. C \textbf{79}, no.6, 474 (2019)



\bibitem{Goncalves:2020vdp}
V.~P.~Gon\c{c}alves, D.~E.~Martins and C.~R.~Sena,
Eur. Phys. J. A \textbf{57}, no.3, 82 (2021)


\bibitem{Goncalves:2020ywm}
V.~P.~Goncalves, D.~E.~Martins and C.~R.~Sena,
Nucl. Phys. A \textbf{1004}, 122055 (2020)


\bibitem{Kharzeev:1998bz}
D.~Kharzeev, H.~Satz, A.~Syamtomov and G.~Zinovjev,
Eur. Phys. J. C \textbf{9}, 459-462 (1999)


\bibitem{Brodsky:2000zc} 
S.~J.~Brodsky, E.~Chudakov, P.~Hoyer and J.~M.~Laget,
Phys.\ Lett.\ B {\bf 498}, 23 (2001)


\bibitem{Redlich:2000cb}
K.~Redlich, H.~Satz and G.~M.~Zinovjev,
Eur. Phys. J. C \textbf{17}, 461-465 (2000)

\bibitem{Frankfurt:2002ka}
L.~Frankfurt and M.~Strikman,
Phys. Rev. D \textbf{66}, 031502 (2002)

\bibitem{Gryniuk:2016mpk} 
O.~Gryniuk and M.~Vanderhaeghen,
Phys.\ Rev.\ D {\bf 94}, no. 7, 074001 (2016)




\bibitem{Hatta:2019lxo}
Y.~Hatta, A.~Rajan and D.~L.~Yang,
Phys. Rev. D \textbf{100}, no.1, 014032 (2019)


\bibitem{Gryniuk:2020mlh} 
O.~Gryniuk, S.~Joosten, Z.~E.~Meziani and M.~Vanderhaeghen,
Phys.\ Rev.\ D {\bf 102}, no. 1, 014016 (2020)



\bibitem{Xu:2020uaa}
Y.~Xu, Y.~Xie, R.~Wang and X.~Chen,
Eur. Phys. J. C \textbf{80}, no.3, 283 (2020)

\bibitem{Zeng:2020coc} 
F.~Zeng, X.~Y.~Wang, L.~Zhang, Y.~P.~Xie, R.~Wang and X.~Chen,
Eur.\ Phys.\ J.\ C {\bf 80}, no. 11, 1027 (2020)

\bibitem{Du:2020bqj}
M.~L.~Du, V.~Baru, F.~K.~Guo, C.~Hanhart, U.~G.~Mei\ss{}ner, A.~Nefediev and I.~Strakovsky,
Eur. Phys. J. C \textbf{80}, no.11, 1053 (2020)

\bibitem{Hatta:2019ocp}
Y.~Hatta, M.~Strikman, J.~Xu and F.~Yuan,
Phys. Lett. B \textbf{803}, 135321 (2020)

\bibitem{Boussarie:2020vmu}
R.~Boussarie and Y.~Hatta,
Phys. Rev. D \textbf{101}, no.11, 114004 (2020)


\bibitem{Mamo:2019mka}
K.~A.~Mamo and I.~Zahed,
Phys. Rev. D \textbf{101}, no.8, 086003 (2020)













\bibitem{Ji:1995sv} 
X.~D.~Ji,
Phys.\ Rev.\ D {\bf 52}, 271 (1995)


\bibitem{Hatta:2018sqd}
Y.~Hatta, A.~Rajan and K.~Tanaka,
JHEP \textbf{12}, 008 (2018)


\bibitem{Lorce:2017xzd}
C.~Lorc\'e,
Eur. Phys. J. C \textbf{78}, no.2, 120 (2018)

\bibitem{Hatta:2018ina}
Y.~Hatta and D.~L.~Yang,
Phys. Rev. D \textbf{98}, no.7, 074003 (2018)



\bibitem{Wang:2019mza}
R.~Wang, J.~Evslin and X.~Chen,
Eur. Phys. J. C \textbf{80}, no.6, 507 (2020)

\bibitem{Metz:2020vxd}
A.~Metz, B.~Pasquini and S.~Rodini,
Phys. Rev. D \textbf{102}, 114042 (2020)



\bibitem{Ji:2021pys}
X.~Ji and Y.~Liu,
[arXiv:2101.04483 [hep-ph]].

\bibitem{Ji:2021mtz}
X.~Ji,
Front.\ Phys.\ (Beijing) {\bf 16}, no. 6, 64601 (2021)

\bibitem{Kharzeev:2021qkd}
D.~E.~Kharzeev,
[arXiv:2102.00110 [hep-ph]].

\bibitem{Wang:2021dis}
R.~Wang, W.~Kou, Y.~P.~Xie and X.~Chen,
Phys.\ Rev.\ D {\bf 103}, no. 9, L091501 (2021)

\bibitem{Kou:2021bez}
W.~Kou, R.~Wang and X.~Chen,
[arXiv:2103.10017 [hep-ph]].




\bibitem{Strakovsky:2019bev}
I.~Strakovsky, D.~Epifanov and L.~Pentchev,
Phys. Rev. C \textbf{101}, no.4, 042201 (2020)


\bibitem{Pentchev:2020kao}
L.~Pentchev and I.~I.~Strakovsky,
Eur. Phys. J. A \textbf{57}, no.2, 56 (2021)


\bibitem{Ali:2019lzf} 
A.~Ali {\it et al.} [GlueX Collaboration],
Phys.\ Rev.\ Lett.\  {\bf 123}, no. 7, 072001 (2019)




\bibitem{vicmiguel2}
V.~P.~Gon\c{c}alves and M.~M.~Jaime,
Phys. Lett. B \textbf{805}, 135447 (2020) 

  \bibitem{Cao:2019gqo}
X.~Cao, F.~K.~Guo, Y.~T.~Liang, J.~J.~Wu, J.~J.~Xie, Y.~P.~Xie, Z.~Yang and B.~S.~Zou,
Phys. Rev. D \textbf{101}, no.7, 074010 (2020)

\bibitem{Xie:2020wfe} 
Y.~P.~Xie, X.~Y.~Wang and X.~Chen,
Chin.\ Phys.\ C {\bf 45}, no. 1, 014107 (2021)
\bibitem{Xie:2020niw} 
Y.~P.~Xie, X.~Cao, Y.~T.~Liang and X.~Chen,
Chin.\ Phys.\ C {\bf 45}, no. 4, 043105 (2021)
\bibitem{Xie:2020ckr} 
Y.~P.~Xie and V.~P.~Goncalves,
Phys.\ Lett.\ B {\bf 814}, 136121 (2021)




\bibitem{EicC}
D.~P.~Anderle, V.~Bertone, X.~Cao, L.~Chang, N.~Chang, G.~Chen, X.~Chen, Z.~Chen, Z.~Cui and L.~Dai, \textit{et al.}
[arXiv:2102.09222 [nucl-ex]].





\bibitem{after} 
  S.~J.~Brodsky, F.~Fleuret, C.~Hadjidakis and J.~P.~Lansberg,
  Phys.\ Rept.\  {\bf 522}, 239 (2013); C.~Hadjidakis {\it et al.},
 Phys. Rept. \textbf{911}, 1-83 (2021)
  
  
\bibitem{Lansberg:2015kha}
J.~P.~Lansberg, L.~Szymanowski and J.~Wagner,
JHEP \textbf{09}, 087 (2015) 

\bibitem{Goncalves:2015hra}
V.~P.~Goncalves and W.~K.~Sauter,
Phys. Rev. D \textbf{91}, no.9, 094014 (2015) 
  
  
   \bibitem{vicmiguel} 
  V.~P.~Goncalves and M.~M.~Jaime,
  Eur.\ Phys.\ J.\ C {\bf 78}, no. 9, 693 (2018)

\bibitem{Lansberg:2018fsy}
J.~P.~Lansberg, L.~Massacrier, L.~Szymanowski and J.~Wagner,
Phys. Lett. B \textbf{793}, 33-40 (2019)




\bibitem{Budnev:1974de}
V.~M.~Budnev, I.~F.~Ginzburg, G.~V.~Meledin and V.~G.~Serbo,
Phys.\ Rept.\  {\bf 15}, 181 (1975)

\bibitem{Lomnitz:2018juf} 
M.~Lomnitz and S.~Klein,
Phys.\ Rev.\ C {\bf 99}, no. 1, 015203 (2019)


\bibitem{Klein:2019avl} 
S.~R.~Klein and Y.~P.~Xie,
Phys.\ Rev.\ C {\bf 100}, no. 2, 024620 (2019)



\bibitem{Klein:2016yzr} 
S.~R.~Klein, J.~Nystrand, J.~Seger, Y.~Gorbunov and J.~Butterworth,
Comput.\ Phys.\ Commun.\  {\bf 212}, 258 (2017)



\bibitem{sakurai} 
  T.~H.~Bauer, R.~D.~Spital, D.~R.~Yennie and F.~M.~Pipkin,
  Rev.\ Mod.\ Phys.\  {\bf 50}, 261 (1978)
  Erratum: [Rev.\ Mod.\ Phys.\  {\bf 51}, 407 (1979)]


\bibitem{glauber} 
  R.~J.~Glauber and G.~Matthiae,
  Nucl.\ Phys.\ B {\bf 21}, 135 (1970)
  



\bibitem{Chekanov:2002xi} 
S.~Chekanov {\it et al.} [ZEUS Collaboration],
Eur.\ Phys.\ J.\ C {\bf 24}, 345 (2002)


\bibitem{Camerini:1975cy} 
U.~Camerini {\it et al.},
Phys.\ Rev.\ Lett.\  {\bf 35}, 483 (1975)


\bibitem{Binkley:1981kv} 
M.~E.~Binkley {\it et al.},
Phys.\ Rev.\ Lett.\  {\bf 48}, 73 (1982)

\bibitem{Frabetti:1993ux} 
P.~L.~Frabetti {\it et al.} [E687 Collaboration],
Phys.\ Lett.\ B {\bf 316}, 197 (1993)

\bibitem{Breitweg:1998ki} 
J.~Breitweg {\it et al.} [ZEUS Collaboration],
Phys.\ Lett.\ B {\bf 437}, 432 (1998)


\bibitem{Adloff:2000vm} 
C.~Adloff {\it et al.} [H1 Collaboration],
Phys.\ Lett.\ B {\bf 483}, 23 (2000)


\bibitem{Chekanov:2009zz} 
S.~Chekanov {\it et al.} [ZEUS Collaboration],
Phys.\ Lett.\ B {\bf 680}, 4 (2009)

\bibitem{Sirunyan:2018sav} 
A.~M.~Sirunyan {\it et al.} [CMS Collaboration],
Eur.\ Phys.\ J.\ C {\bf 79}, no. 3, 277 (2019)




















\bibitem{lhcfixed} 
  E.~Maurice [LHCb Collaboration],
  arXiv:1708.05184 [hep-ex].


\bibitem{smog} 
  R.~Aaij {\it et al.} [LHCb Collaboration],
  JINST {\bf 9}, no. 12, P12005 (2014)

\bibitem{Aaij:2018ogq}
R.~Aaij \textit{et al.} [LHCb],
Phys. Rev. Lett. \textbf{122}, no.13, 132002 (2019)
  
\bibitem{Maciula:2020cfy}
R.~Maciu\l{}a,
Phys. Rev. D \textbf{102}, no.1, 014028 (2020)  
  
  
\bibitem{antiproton} 
  M.~Korsmeier, F.~Donato and M.~Di Mauro,
Phys. Rev. D \textbf{97}, no.10, 103019 (2018)







\end{thebibliography}
\end{document}